**Spin Seebeck effect at low temperatures in the nominally paramagnetic insulating state of vanadium dioxide**

Renjie Luo[1], Xuanhan Zhao[1], Liyang Chen[2], Tanner J. Legvold[1], Henry Navarro[3], Ivan K. Schuller[3], Douglas Natelson[1,4*]

[1]Department of Physics and Astronomy, Rice University, Houston TX 77005, USA

[2]Applied Physics Graduate Program, Rice University, Houston TX 77005, USA

[3]Department of Physics and Center for Advanced Nanoscience, University of California-San Diego, La Jolla CA 92093, USA

[4]Department of Electrical and Computer Engineering and Department of Materials Science and NanoEngineering, Rice University, Houston TX 77005, USA

**The low temperature monoclinic, insulating phase of vanadium dioxide is ordinarily considered nonmagnetic, with dimerized vanadium atoms forming spin singlets, though paramagnetic response is seen at low temperatures. We find a nonlocal spin Seebeck signal in $VO_2$ films that appears below 30 K and which increases with decreasing temperature. The spin Seebeck response has a non-hysteretic dependence on in-plane external magnetic field. This paramagnetic spin Seebeck response is discussed in terms of prior findings on paramagnetic spin Seebeck effects and expected magnetic excitations of the monoclinic ground state.**

Vanadium dioxide is an archetypal strongly correlated transition metal oxide, with a phase transition at ~345 K in bulk between a high temperature, rutile metallic phase with 1D vanadium chains and a low temperature, monoclinic insulating phase with dimerized vanadium atoms. The vanadium dimers are thought to form singlets, greatly reducing the magnetic response of the $VO_2$ at low temperatures[1-3].

---

* Corresponding author: natelson@rice.edu

The spin Seebeck effect (SSE) has proven useful in characterizing angular momentum transport in magnetic insulators[4, 5]. In measurements of the nonlocal spin Seebeck effect (nlSSE), the current flowing through a heater wire on the surface of a material of interest is driven at angular frequency $\omega$, leading to a temperature gradient with a DC offset and an AC component at $2\omega$. A properly oriented angular momentum current is driven by the temperature gradient (e.g. a flux of magnons in a ferrimagnet[6] or antiferromagnet[7]) and a voltage at $2\omega$ is detected at nearby inverse spin Hall detector made from a strong spin-orbit metal (e.g. Pt). The nlSSE was first observed in the ferrimagnetic insulator $Y_3Fe_5O_{12}$ (YIG), where the magnon spin diffusion length is found to be several micrometers at room temperature[8]. This nonlocal approach to examining the SSE has been applied to different magnetic systems, including antiferromagnets (NiO[7], $\alpha$-$Fe_2O_3$[9], $\alpha$-$Cr_2O_3$[10]). The local SSE has been observed in paramagnets ($Gd_3Ga_5O_{12} \equiv$ GGG, $DyScO_3$[11]; $La_2NiMnO_6$ above its ferromagnetic ordering temperature[12]; the quantum spin chain compound $Sr_2CuO_3$[13]). In GGG, while there is no long-range magnetic order, there is short-range order[14] and field-induced long-range order that are thought to play a role in the transport of angular momentum as detected by the local SSE[15]. Thus far there are no reports of the nlSSE in paramagnets.

In this work we report a nonlocal spin Seebeck response in $VO_2$ films at low temperatures deep in the insulating regime. Given the expected nature of paramagnetism in this material, this observation is surprising. For a fully spin dimerized system, paramagnetic response is expected to be suppressed at low temperatures, unlike, e.g., a classical spin liquid (such as GGG). The "local" paramagnetism, as in the van Vleck mechanism, is expected to be temperature independent, not expected to transport angular momentum and thus should not contribute to the SSE response. The SSE is non-hysteretic as a function of **B** and has the expected dependence on orientation of magnetic field in-plane. With the field in-plane, the SSE signal grows in magnitude with decreasing temperature below an onset of detectability at approximately 30 K, with a sign inversion between 15 K and 10 K.

The ~100 nm thick epitaxial $VO_2$ film was grown by reactive sputtering on top of an $Al_2O_3$ substrate (r-cut). A 4-mtorr argon/oxygen mix (8% $O_2$) was used during deposition, and the substrate was kept at 520 °C during the growth and later cooled down at a rate of 12 °C min$^{-1}$. X-ray diffraction measurements confirmed single-phase growth, textured along (100) for $VO_2$. Transport measurements were carried out on a TTPX Lakeshore cryogenic probe station, using a Keithley 6221 current source and a Keithley 2182A nanovoltmeter. Electron beam lithography and magnetron sputtering are used to prepare Pt nanowires (100 μm long, 20 nm thick, 200 nm wide, separated by 400 nm) on the $VO_2$ surface for the transverse SSE measurements, shown in Fig. 1b. Typical resistance of each wire is 18-20 kΩ. The films exhibit a large, hysteretic metal-insulator transition (Fig. 1c), as expected for high quality films. Measurements are performed as a function of temperature and field in a Quantum Design Physical Property Measurement System equipped with a rotation stage.

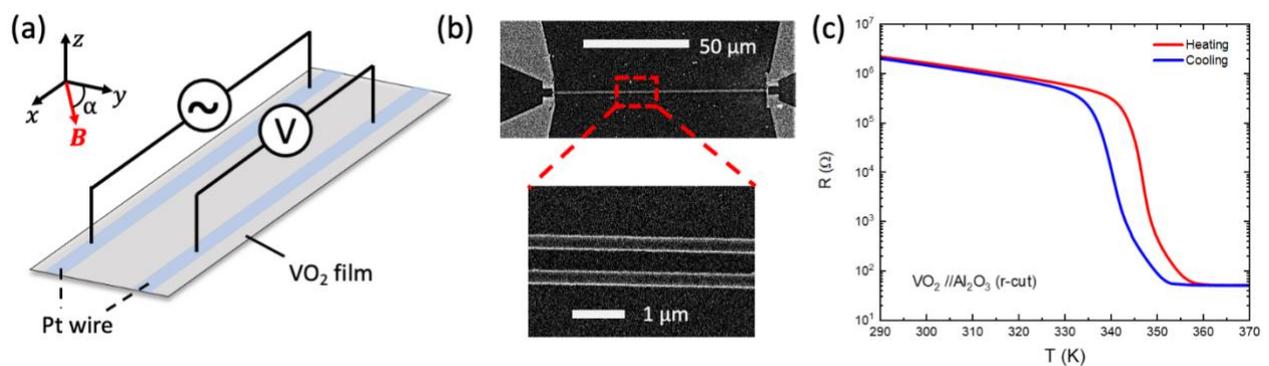

FIG. 1. (a) Geometry of the transverse spin Seebeck measurement configuration, showing the field orientations in the *x-y* plane (*α*). (b) Scanning electron micrograph showing a typical device configuration. (c) Resistance as a function of temperature for a typical $VO_2$ films used in this work, showing the metal-insulator transition.

As shown, an AC heater current at angular frequency $\omega = 2\pi \times (7.7\text{ Hz})$ is driven through a Pt wire, while the voltage across the other wire is measured at $\omega$ and $2\omega$ using a lock-in amplifier. The overwhelmingly dominant signal at $\omega$ is due to capacitive coupling between the wires. Even in the absence of $VO_2$ (*e.g.* Pt wires fabricated on $SiO_2$/Si or sapphire substrates) there is a temperature-dependent, magnetic field-independent parasitic background signal at $2\omega$ in this electrode and wiring geometry (see Supplemental Material Fig S2).

Below 30 K, a magnetic field-dependent $2\omega$ signal becomes detectable, as shown in Fig. 2a, b for in-plane field oriented along $\alpha = 0°$. The signal magnitude increases with increasing field, linearly near $\boldsymbol{B} = 0$ T and saturates at high fields, with no indication of hysteresis as a function of $\boldsymbol{B}$. As the temperature decreases, the magnitude of the signal increases, with a change in sign between 15 K and 10 K. The signal grows rapidly with further decreases in temperature, and the saturation at large $\boldsymbol{B}$ is readily apparent.

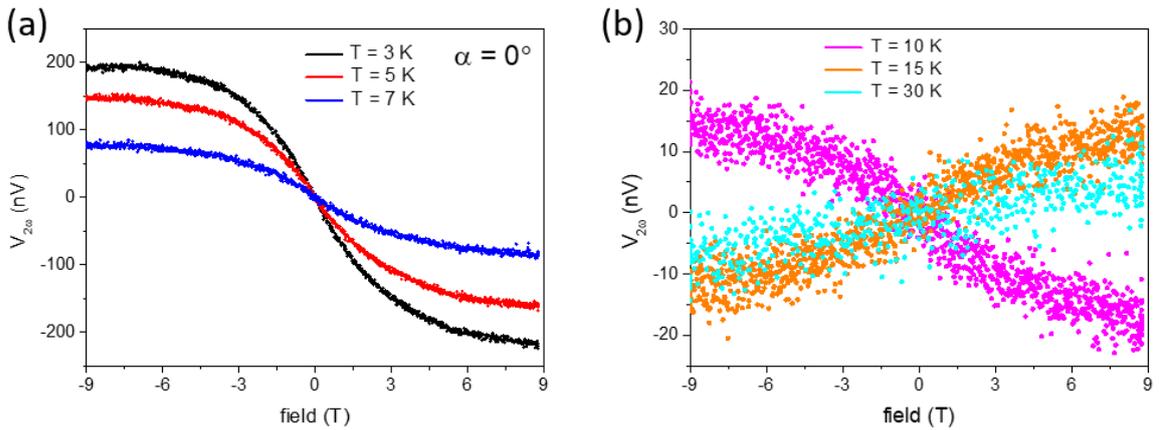

FIG 2. (a) The second harmonic signal vs. $\boldsymbol{B}$ at low temperatures for $\alpha = 0°$. (b) The second harmonic signal at higher temperatures, showing the sign inversion between 10 K and 15 K for the measurement configuration in Fig. 1b. The heating power is 40 µW.

To confirm that this $2\omega$ signal results from the spin Seebeck effect, it is important to consider the field dependence of the signal on the orientation of $\boldsymbol{B}$ in the plane of the film. As shown in Fig. 3a, at each temperature and fixed field $B$ = 3 T, the signal is fit well by a cos $\alpha$ dependence, as expected for the SSE and required for the generation of an inverse spin Hall signal in the Pt associated with transport of angular momentum from magnetization oriented along the *y*-axis in the geometry of Fig. 1a. The 2$\omega$ voltage signal likewise depends linearly on heater power at fixed $B$ oriented at $\alpha$ = 0°, as expected for a signal of spin Seebeck origin (Fig. 3b). A possible confounding effect could be the Nernst response of Pt, but that is expected to be temperature independent[11] and would require a vertical temperature gradient at the VO$_2$/Pt detector interface. Instead, the temperature dependence of the signal is qualitatively similar to the paramagnetic response reported in the susceptibility[3, 16], supporting that the measured signal originates from the SSE. In addition to Nernst response, another possible contribution to a 2$\omega$ signal in the detector would be local SSE at the detector due to a local temperature gradient[17]. However, our temperature profile simulations (see SI) show that the vertical temperature difference between detector Pt and underlying VO$_2$ film is on the order of $10^{-4}$ mK (Fig. S9), far too small to lead to any measurable Nernst or local spin Seebeck effects.

Extensive additional control experiments are shown in the Supplemental Material. Concerns about charge leakage, unintended capacitive couplings, and other artifacts have been raised in nonlocal spin transport experiments in conducting samples[18]. We examined the 2$\omega$ signal present in the identical Pt electrode geometry, fabricated directly on a sapphire substrate and find no magnetic field dependence for in-plane fields (Fig. S1). We performed a temperature-dependent study of this background as a function of wiring configuration on a SiO$_2$/Si substrate (Fig. S2), and again find that it is field independent. Measurements on VO$_2$ devices with a Au wire and a Pt wire detector wire show a field-dependent 2$\omega$ SSE response when the Pt wire is used as the detector, and no temperature/field-dependent response when the Au wire is used as the detector (Fig. S5). Possible charge leakage between heater and detector is expected to lead to a 1$\omega$ signal rather than a 2$\omega$ signal; the 1$\omega$ signal is dominated by capacitive effects (out of phase response larger than the in-phase response) and shows no clear field dependence at 5 K (Fig. S6). These controls, as well as the systematic dependence of the measured signal on temperature, power, field magnitude and direction, and device geometry are consistent with nlSSE response as the signal origin.

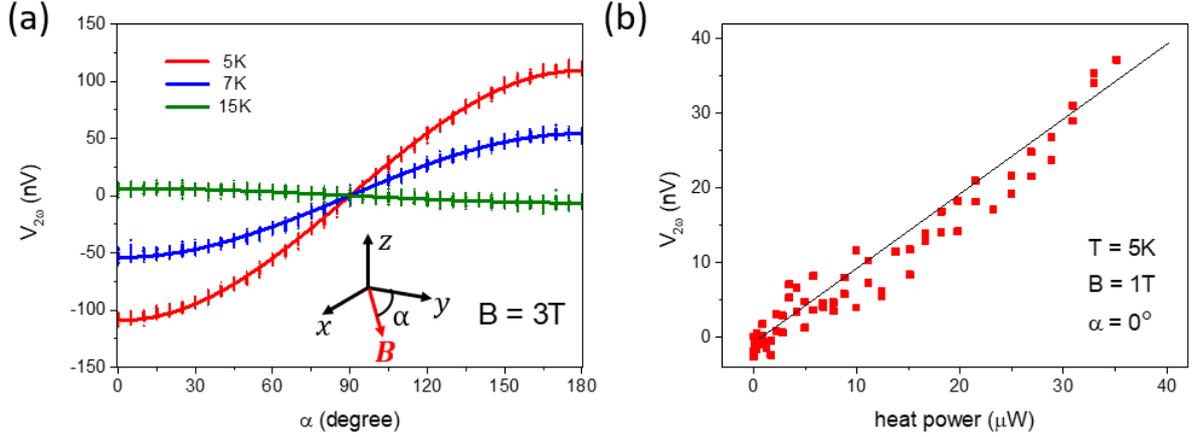

FIG. 3. (a) The second harmonic signal as a function of in-plane field orientation. Vertical marks show the scatter in the data at each angle, and solid lines are fits to the $\cos\alpha$ dependence. (b) The second harmonic signal is linear in heater power, as expected.

The growth of the SSE signal magnitude with decreasing temperature is consistent with observations of local SSE response in other paramagnets[11]. In these experiments the longitudinal SSE response as a function of $B$ resembles the experimentally determined $M$ vs $B$ polarization. Deviations in the SSE vs $B$ in GGG from a simple Brillouin function-like dependence become apparent below 4 K and are thought to be related to short-range magnetic order[15]. Because of the small volume of $VO_2$ material relative to the volume of sapphire substrate, and paramagnetic impurities contained within the sapphire[19], it has not been possible to isolate directly $M$ vs $B$ for the $VO_2$ films in our samples. The low temperature magnetization $M$ in bulk $VO_2$ crystals [3] does not saturate at high fields.

The sign reversal in the measured SSE between 10 K and 15 K is qualitatively similar to a sign reversal observed in nlSSE measurements on YIG films[6]. In this system, spin transport takes place via magnons, with regions of elevated magnon chemical potential building up in proximity to the injecting and collecting Pt wires[20]. The SSE sign reversal in that case is a consequence of a crossover of length scales between Pt electrode spacing and magnon chemical potential spatial scale, affected strongly by the thickness of the magnon-bearing YIG film on the GGG substrate. In the present experiment, the sign reversal in SSE response moves to lower temperatures with increasing electrode spacing (see Supplemental Material Fig S3), though a detailed study has not been performed. Sign reversal of the SSE response as a function of temperature can also take place due to other mechanisms. In antiferromagnets and compensated ferrimagnets, competing responses of two magnon branches can cause a crossover in SSE sign as a function of temperature[21]. Similarly, in an antiferromagnet, the presence of multiple magnon branches can lead to a field-dependent precession of the magnon pseudospin[22]. Both of these possibilities would require the presence of long-range magnetic order and multiple magnon branches in the material, which are believed absent in $VO_2$.

Repeated measurements of devices, after cycling back to room temperature and sample aging on the timescale of days or weeks, show much reduced signal magnitudes, though identical field dependences. The nlSSE signal depends critically on interfacial exchange coupling between the oxide and the Pt conduction electrons, and hence is extremely sensitive to the Pt/magnetic insulator interface quality. The observed signal reduction upon sample aging likely indicates degradation of the Pt/$VO_2$ interface.

The presence of a strong spin Seebeck response that grows at low temperatures in this material calls into question the nature of the angular momentum-carrying excitations. In prior examinations of paramagnets, the argument was advanced that SSE response can originate from short-range magnetic order[15] or field-supported paramagnons at temperatures above a low ordering temperature[11]. The magnetic state of monoclinic VO$_2$ is thought to be a singlet-dimer state[1], rather than a traditionally ordered state. One picture for the phase is that the dominant spin-carrying excitations are thermally excited triplets ("triplons") and the increase in susceptibility at low temperatures in the material is due to the Curie-like response of the temperature-dependent effective spin of the dimers[3, 16]. This picture does not have long-ranged order nor well-defined magnons. The sign inversion of the SSE as a function of temperature suggests that it may be related to the spatial distribution of the local chemical potential of spin-carrying excitations.

A triplon SSE has been reported in the local measurement configuration in the spin-Peierls system CuGeO$_3$[23], where Cu atoms form one-dimensional spin-1/2 chains with antiferromagnetic exchange interactions. The most important feature of the triplon SSE is that it has the opposite sign of voltage to the magnon SSE in ferromagnetic insulators. This is because triplon spin current is carried by an excitation with spin direction parallel to the external field, while magnons carry angular momentum antiparallel to the field. Another feature of the triplon local SSE is that its magnitude does not scale with the $M(H)$ response and does not saturate at high fields, similar to what is observed in the VO$_2$ nlSSE. Based on this phenomenology, VO$_2$ is another candidate for the observation of the triplon SSE, though the sign reversal as a function of temperature in the nlSSE measurement complicates interpretation.

We note one report in the literature[24] in which muon spin rotation implies the onset of an internal magnetic field within monoclinic VO$_2$ below a sharp transition near 35 K. This is interpreted as originating from the "disruption of the V-V dimers" which produce a "non-zero net spin" below 35 K. This suggests that there may indeed be some magnetic ordering taking place that is not readily observed in the susceptibility yet may lead to well-defined magnon excitations. While high temperature defect-mediated ferromagnetism is possible in $VO_2$ films deposited on sapphire[25], there is no evidence of any ferromagnetic order in the present or other materials grown by the authors, nor is the observed field dependence compatible with ferromagnetism.

To summarize, we have detected a spin Seebeck response in insulating $VO_2$ films below 30 K that grows in magnitude with decreasing temperature. This response is nonhysteretic in **B**, has the expected SSE dependence on field orientation in the plane, and saturates at large **B** and low temperatures. Given the singlet-dimer nature expected of the monoclinic $VO_2$ ground state, a detailed examination of angular momentum transport in this system is needed to explain these observations, and whether such an effect is a consequence of the transport of triplet excitations or a sign of the emergence of magnon-like excitations in a magnetically ordered phase below 35 K.

*Supplementary Materials*

See supplementary materials for additional control experiments and thermal modeling.

*Acknowledgments*

The authors acknowledge Shusen Liao for his assistance with finite element modeling. XZ, DN, RL and TJL acknowledge support from DMR-1704264 and DMR-2102028 for spin Seebeck measurements. DN and LC acknowledge support from DOE BES award DE-FG02-06ER46337 for data acquisition and development for spin transport in $VO_2$. Synthesis, characterization, joint design of the experiments, extensive discussions and joint writing of the manuscript were funded by the U.S. Department of Energy, Office of Science, Basic Energy Sciences under Award # DE-FG02-87ER45332.

*Authors Declarations*

*Conflict of Interest*

The authors have no conflicts to disclose.

*Data Availability*

The data that support the findings of this study are available from the corresponding author upon reasonable request.

**Supplementary Materials**

**Spin Seebeck effect at low temperatures in the nominally paramagnetic insulating state of vanadium dioxide**

Renjie Luo[1], Xuanhan Zhao[1], Liyang Chen[2], Tanner J. Legvold[1], Henry Navarro[3], Ivan K. Schuller[3], Douglas Natelson[1,4]

[1]*Department of Physics and Astronomy, Rice University, Houston TX 77005, USA*

[2]*Applied Physics Graduate Program, Rice University, Houston TX 77005, USA*

[3]*Department of Physics and Center for Advanced Nanoscience, University of California-San Diego, La Jolla CA 92093, USA*

[4]*Department of Electrical and Computer Engineering and Department of Materials Science and NanoEngineering, Rice University, Houston TX 77005, USA*

**1. Control experiments on sapphire substrates.**

An analogous device with the same Pt electrode sizes was patterned on a single-crystal sapphire substrate by the same fabrication methods. The $2\omega$ signals measured on it show trivial behavior at both high and low temperature, with the in-plane external field perpendicular to the Pt wire (Fig. S1). This indicates that sapphire itself cannot support a spin Seebeck effect (SSE) or Nernst response, as expected. This confirms that the SSE observed on VO$_2$/sapphire devices originates from the VO$_2$ film. Likewise, this demonstrates that there is no inherent crosstalk or leakage in the measurement system itself that could lead to spurious $2\omega$ signals.

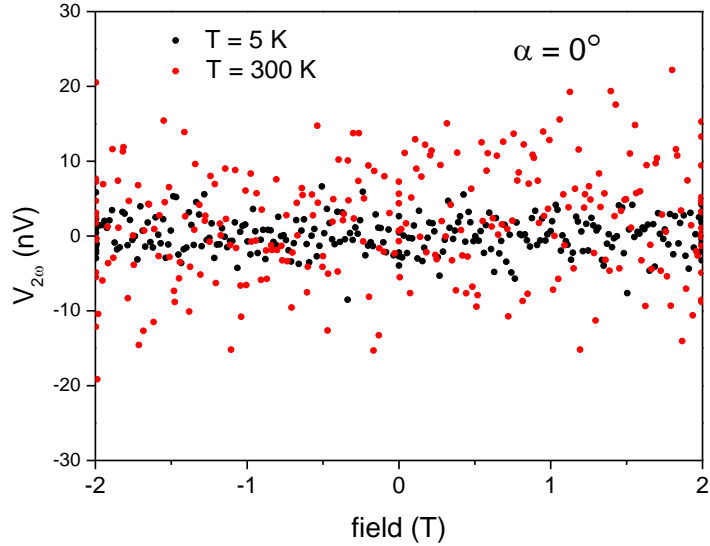

**FIG. S1.** Field dependence of the second harmonic signal at low and high temperatures for α = 0° on a device patterned on a sapphire substrate.

## 2. Zero-field 2ω background offset with different wiring configuration on SiO$_2$/Si substrates.

When measuring the spin Seebeck signal, there is a background response at 2ω as detected by the lock-in amplifier (the signal sitting at the zero field, or zero-field offset) that changes slightly with temperature. To understand this, we patterned analogous devices, with the same size as the Pt wires for the VO$_2$ measurements, instead on SiO$_2$/Si substrates by the same fabrication methods. With different electrode and wiring configurations, we found that the zero-field offset either increases or decreases in the temperature range from 5 K to 20 K and remains essentially constant at temperatures above 20 K (Fig. S2a). The value of the zero-field offset changes greatly when changing the wiring and grounding configuration. This implies that this offset originates from the capacitive and inductive couplings between the wires in the measurement apparatus and the device. No field dependence of this background offset is observed in devices on SiO$_2$/Si, and hence it has been subtracted from the VO$_2$ data, under the presumption that its origin is extrinsic to the sample and its field independence will remain.

For the data shown in Fig. S2a, "normal" refers to a configuration with heater current applied to one wire, AC voltage measured on the other wire, and the ground of the drive current is on the same end of the wires as the "B" of the (A-B) differential voltage measurement. The "swap" configuration switches which wire is used as the heater and which is used as the detector. The "reverse" configuration has the ground of the drive current on the opposite end of the wires as the B contact of the differential voltage measurement.

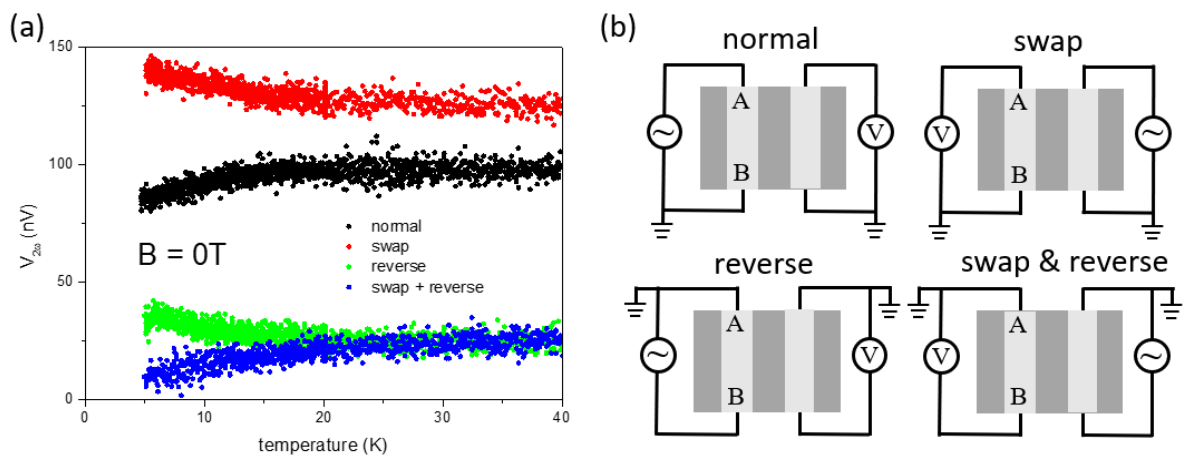

FIG. S2. (a) Temperature dependence of the zero-field offset with different wiring configurations on SiO$_2$/Si substrates. (b) Illustration of different electrode and wiring configurations.

## 3. Spin Seebeck signal with different separations.

Devices with the different wire separations were patterned on a different piece of VO$_2$ film/sapphire substrate other than the device reported in the main text and the spin Seebeck signal was measured at $\alpha = 0°$ and low temperatures. For devices with wire separation 0.4 μm, the near-zero-field slope is negative at 5 K and 10 K, and becomes positive at 15 K (Fig. S3a). However, for device with separation 1 μm, the near-zero-field slope changes sign between 7 K and 10 K (Fig. S3b). We note that for this set of devices (co-fabricated simultaneously), the low temperature SSE voltage vs field appears more complicated than the saturating response at high fields shown in the main manuscript, and the magnitude of the signal was lower, suggesting comparatively inferior Pt/VO$_2$ interface quality.

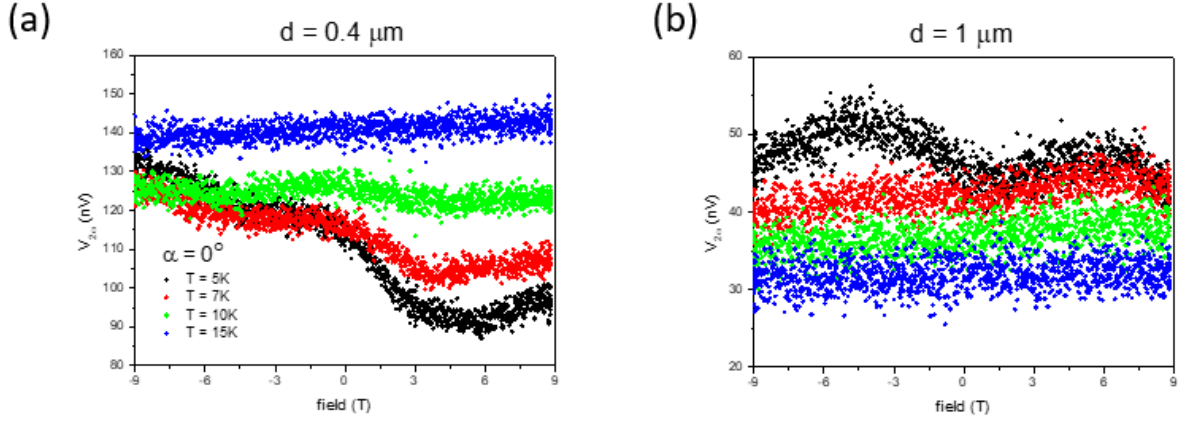

FIG. S3. Field dependence of the second harmonic signal at different temperatures with separation between the wires being 0.4 μm (a) and 1 μm (b).

## 4. Resistance and capacitance between two nearby Pt wires.

We used two-point lock-in methods to measure the resistance and capacitance between two nearby Pt wires, with 0.4 μm separation on top of the VO$_2$ film on sapphire. An ac voltage with frequency $f$=2.01Hz is applied to the two Pt wires, and the voltage $U$ between the wires and the current flowing through the VO$_2$ film are measured simultaneously. The current has both x-component and y-component because of the capacitance between the wires, denoted as $I_x$ and $I_y$. Then the resistance of the film is given by: $R = U/I_x$ and the capacitance is given by $C = I_y/(2\pi f U)$.

Although $VO_2$ is deep in the insulating regime, the resistance between two nearby Pt wires is not unmeasurably high in this configuration. This residual conduction has no obvious temperature dependence from 5 K to 30 K (Fig. S4b). Moreover, the capacitance between the wires is also independent of the temperature (Fig. S4c).

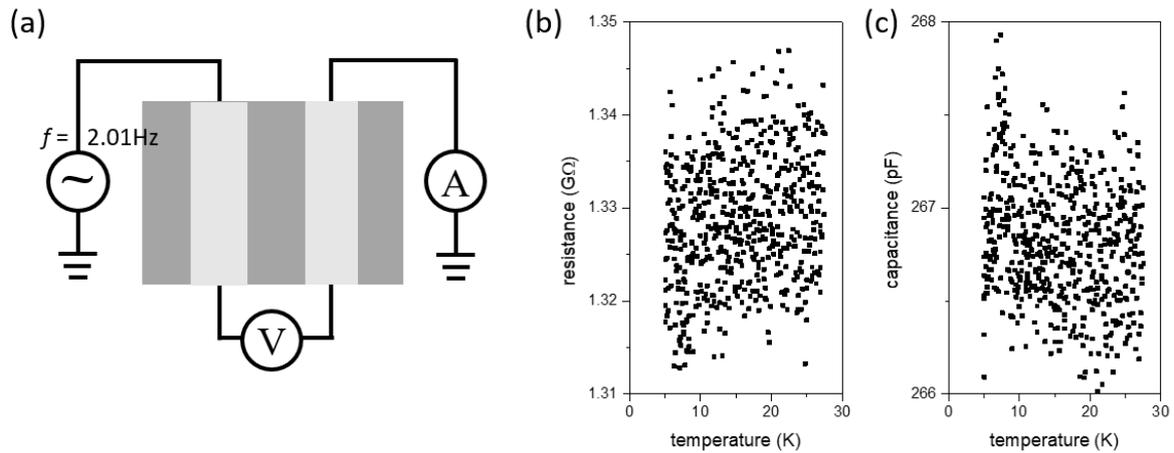

FIG. S4. (a) Illustration of two-point lock-in measurement setup. (b, c) Temperature dependence of resistance and capacitance between two nearby Pt wires measured by two-point lock-in technique.

## 5. Control experiment with Au wire on $VO_2$/sapphire substrates.

An similar device with the same electrode sizes was patterned on the $VO_2$ film/sapphire substrate by the same fabrication methods, one wire being Pt and one wire being Au. In this device, the resistance of Pt wire is ~18 k$\Omega$ and the resistance of Au wire is ~600 $\Omega$ at 10K. When using Au wire as heater and Pt wire as detector, the $2\omega$ signals measured on Pt wire show the same behavior (Fig. S5a) as that in the main text. However, when using Pt wire as heater and measuring $2\omega$ voltage along Au wire under the same heating power condition, it shows trivial behavior (Fig. S5b), since Au has comparatively small spin Hall angle. This confirms that the SSE observed on VO$_2$/sapphire devices should originate from the Pt.

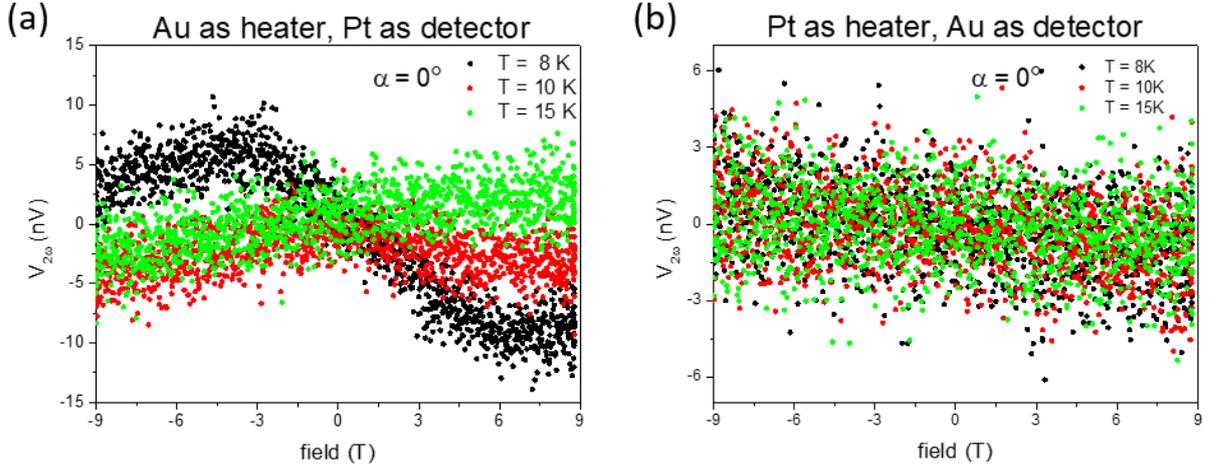

FIG. S5. Field dependence of the 2ω signal at different temperatures for α = 0° on a device patterned on a VO$_2$/sapphire substrate with Pt wire (a) or Au wire (b) playing role of detector. The heating power is 40 μW for both cases.

**6. First harmonic signal on VO$_2$/sapphire substrates.**

The first harmonic signal across the detector of the device on VO2/sapphire substrates is has no clear trend with field, as shown in the Fig. S6. The *x*-channel voltage is relatively small, given that the voltage applied on the heater is ~0.8V, indicating again the insulating nature of VO2 at low temperatures. The *y*-channel voltage is dominant and is expected to arise from the capacitance between two nearby Pt wires. The small size of the field dependence (consistent with drift during the measurement) is readily apparent through the in the data in Fig. S6. There is no obvious signature of direct spin injection and detection as has been seen in other nlSSE measurements in, e.g., YIG. This may be due to poor efficiency of injection due to Pt/VO$_2$ interface quality, or due to a short lifetime or poor mobility of spin-carrying excitations.

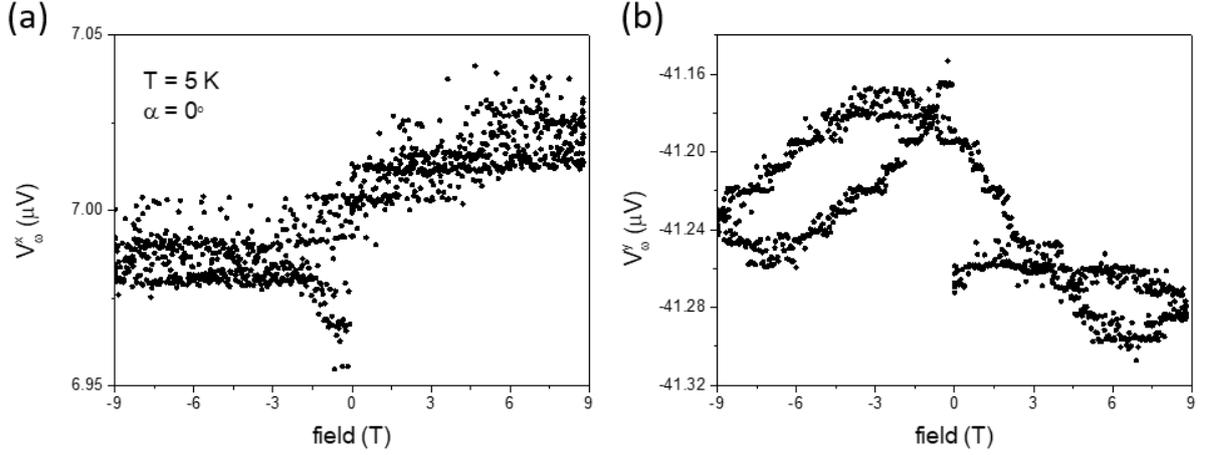

FIG S6. Field dependence of the ω signal at 5K for α = 0° on a device patterned on a VO$_2$/sapphire substrate with *x*-channel (a) and *y*-channel (b). There is no clear signature of direct spin transport or charge leakage in these data.

## 7. Temperature profile simulation of nlSSE device on VO$_2$/sapphire substrates.

To address concerns about possible confounding signals at the detector (e.g. Nernst or local SSE response due to a temperature gradient between the Pt detector and the immediately underlying VO$_2$), we calculated the temperature profiles (using the finite element software COMSOL) of the nlSSE device induced by Joule heating on VO$_2$/sapphire substrates with the specific sample geometry and reasonable material parameters (see Table S1). The PPMS chamber temperature $T_0 = 5$ K is set as a boundary condition at the bottom of sapphire. We also assume a thermal boundary conductance at the Pt/VO$_2$ interface (~ 60 MW m$^{-2}$T$^{-1}$), an order-of-magnitude estimate based on interfacial thermal transport measurements in SSE devices[S1]. As seen in Fig. S7, the current flowing through the heater wire with power 32 μW warms up the heater and underlying VO$_2$ nearby; however, this temperature gradient is strongly localized near the heater wire and the heating has only a very minor effect on the detector wire. More specifically, the temperature difference between the detector and the underlying VO$_2$ film is on the order of 10$^{-4}$ mK. This demonstrate that for reasonable values of all parameters, there is not a detectable contribution to the measured signal from vertical thermal gradients at the detector wire.

|  | Pt | $VO_2$ | sapphire |
|---|---|---|---|
| geometry | length 100 μm; width 200 nm; thickness 20 nm; edge-to-edge distance 400 nm. | Infinite plane; thickness 100 nm. | Infinite plane; infinite thickness |
| Electric conductivity | $1.25 \times 10^6$ S/m | ~ $10^{-5}$ S/m Insulating state | ~ $10^{-10}$ S/m |
| thermal conductivity | ~ 0.15 W/(mK) | ~ 20 W/(mK) | ~ 550 W/(mK) |
| density | $21.45 \times 10^3$ kg/m$^3$ | $4.57 \times 10^3$ kg/m$^3$ | $3.98 \times 10^3$ kg/m$^3$ |
| specifit heat | ~ 0.266 J/(kgK) | ~ 0.0313 J/(kgK) | ~ 0.098 J/(kgK) |

TABLE S1. Geometry and material parameters used in the finite element simulation.

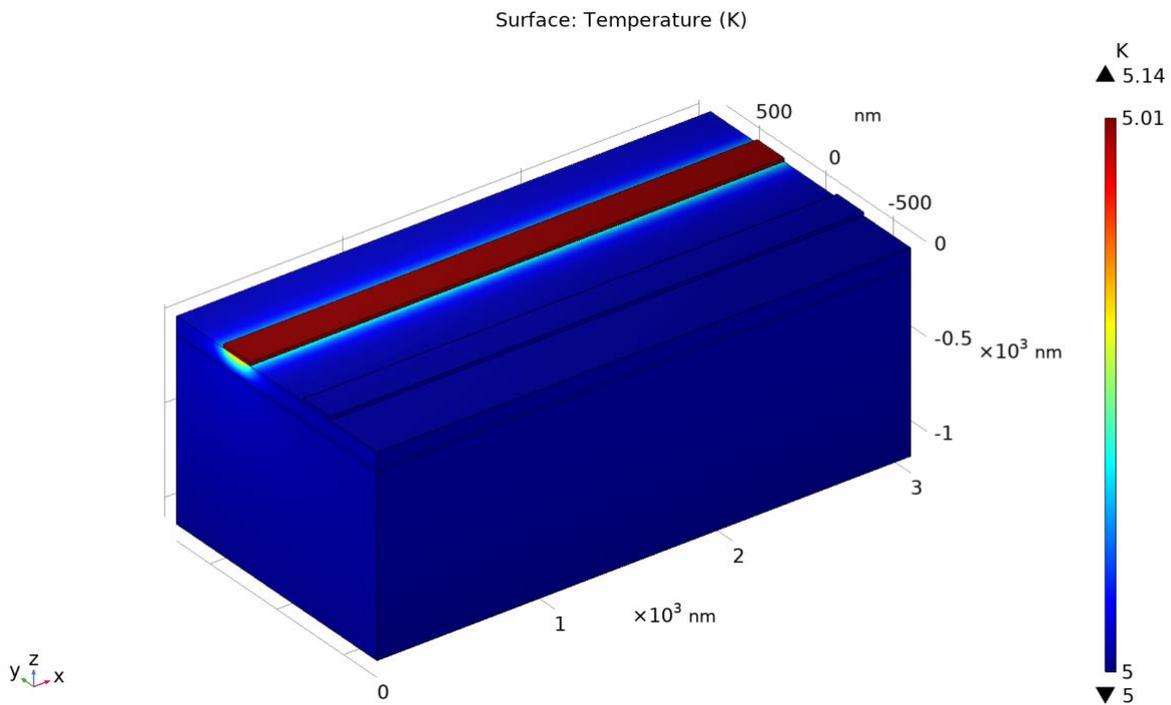

FIG S7. Temperature profile of the nlSSE device on the $VO_2$/sapphire substrates.

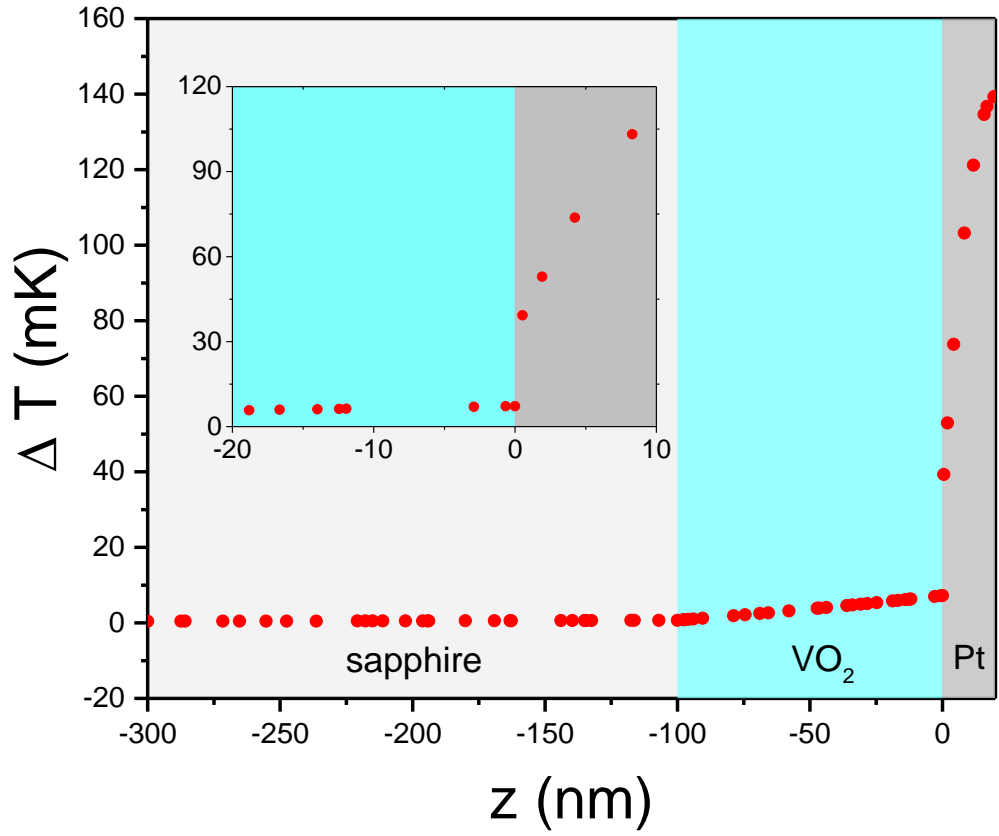

FIG S8. Temperature difference on the z-direction at the center of the heater wire and underlying VO2/sapphire. Inset is the enlargement of the figure near the Pt/VO$_2$ interface. Light gray, cyan and gray color indicates the sapphire, VO$_2$, and platinum, respectively.

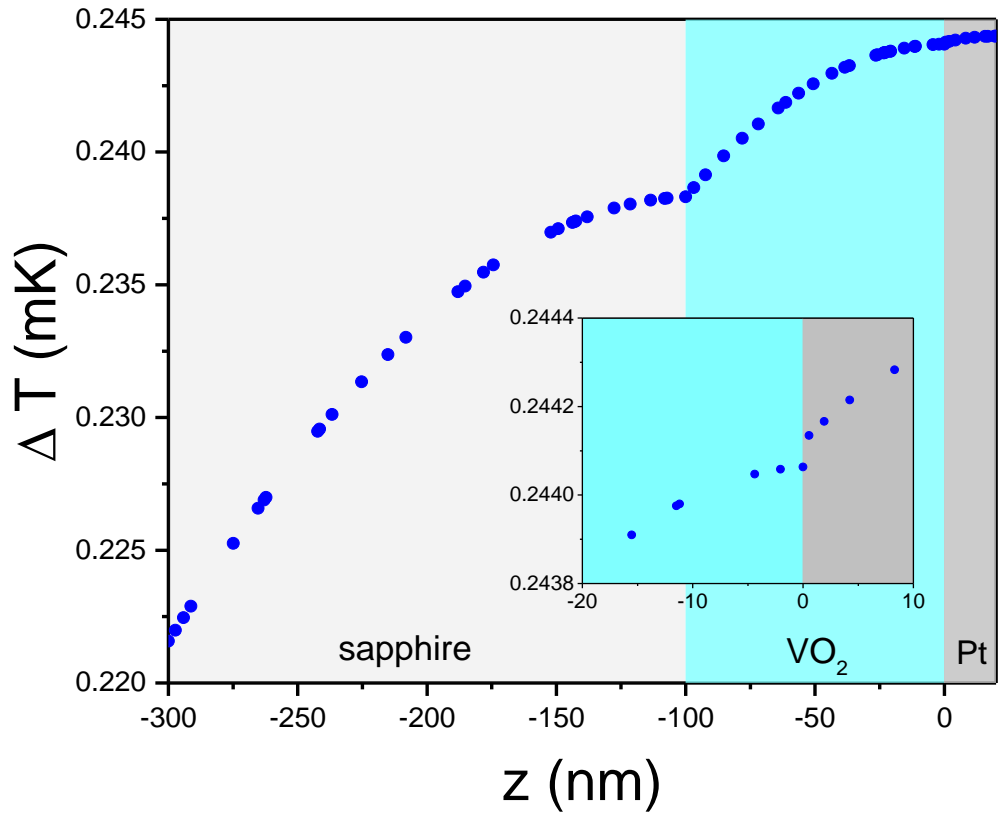

FIG S9. Temperature difference on the z-direction at the center of the detector wire and underlying VO2/sapphire. Inset is the enlargement of the figure near the Pt/VO$_2$ interface.

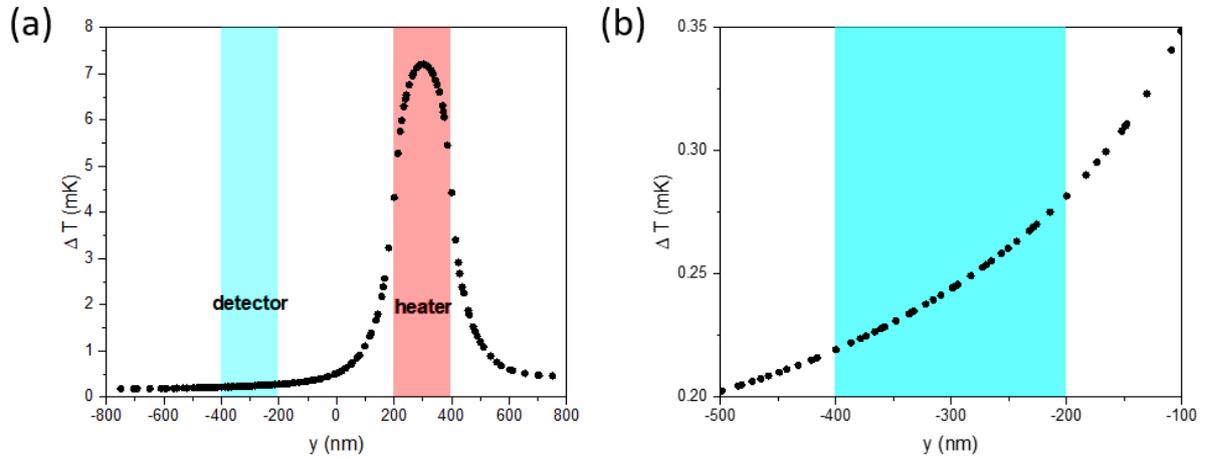

FIG S10. (a) Temperature difference on the y-direction on the top surface of $VO_2$, including the interface between $VO_2$ film and the heater wire and the interface between $VO_2$ film and the detector wire. (b) Temperature difference on the y-direction on the top surface of $VO_2$ near the interface between $VO_2$ film and the detector wire.

S1. Frank Angeles, Qiyang Sun, Victor H. Ortiz, Jing Shi, Chen Li, and Richard B. Wilson, Phys. Rev. Materials **5**, 114403 (2021).